\documentclass[sigconf, nonacm=true]{acmart}

\usepackage{multirow}
\usepackage{algpseudocode}
\usepackage{algorithm}
\usepackage{graphicx}
\usepackage{subcaption}
\usepackage{array}
\usepackage[export]{adjustbox}
\usepackage{flushend}

\makeatletter
\g@addto@macro{\UrlBreaks}{\UrlOrds}
\makeatother

\begin{document}
\title{Visualizing Webpage Changes Over Time}

\author{Abigail Mabe}
\affiliation{
  \department{Department of Computer Science}
  \institution{Old Dominion University}
  \city{Norfolk}
  \state{VA}
  \postcode{23529}
  \country{USA}
}
\email{amabe002@odu.edu}

\author{Dhruv Patel}
\affiliation{
  \department{Department of Computer Science}
  \institution{Old Dominion University}
  \city{Norfolk}
  \state{VA}
  \postcode{23529}
  \country{USA}
}
\email{dpate006@odu.edu}

\author{Maheedhar Gunnam}
\affiliation{
  \department{Department of Computer Science}
  \institution{Old Dominion University}
  \city{Norfolk}
  \state{VA}
  \postcode{23529}
  \country{USA}
}
\email{mgunn001@odu.edu}

\author{Surbhi Shankar}
\affiliation{
  \department{Department of Computer Science}
  \institution{Old Dominion University}
  \city{Norfolk}
  \state{VA}
  \postcode{23529}
  \country{USA}
}
\email{sshan001@odu.edu}

\author{Mat Kelly}
\orcid{0000-0002-0236-7389}
\affiliation{%
  \department{College of Computing \& Informatics}
  \institution{Drexel University}
  \city{Philadelphia}
  \state{PA}
  \postcode{19104}
  \country{USA}
}
\email{mkelly@drexel.edu}

\author{Sawood Alam}
\orcid{0000-0002-8267-3326}
\affiliation{%
  \department{Department of Computer Science}
  \institution{Old Dominion University}
  \city{Norfolk}
  \state{VA}
  \postcode{23529}
  \country{USA}
}
\email{salam@cs.odu.edu}

\author{Michael L. Nelson}
\orcid{0000-0003-3749-8116}
\affiliation{%
  \department{Department of Computer Science}
  \institution{Old Dominion University}
  \city{Norfolk}
  \state{VA}
  \postcode{23529}
  \country{USA}
}
\email{mln@cs.odu.edu}

\author{Michele C. Weigle}
\orcid{0000-0002-2787-7166}
\affiliation{%
  \department{Department of Computer Science}
  \institution{Old Dominion University}
  \city{Norfolk}
  \state{VA}
  \postcode{23529}
  \country{USA}
}
\email{mweigle@cs.odu.edu}

\begin{abstract}
We report on the development of TMVis, a web service to provide visualizations of how individual webpages have changed over time. We leverage past research on summarizing collections of webpages with thumbnail-sized screenshots and on choosing a small number of representative past archived webpages from a large collection. We offer four visualizations: image grid, image slider, timeline, and animated GIF. Embed codes for the image grid and image slider can be produced to include these on separate webpages. The animated GIF can be downloaded as an image file for the same purpose. This tool can be used to allow scholars from various disciplines, as well as the general public, to explore the temporal nature of web archives. We hope that these visualizations will just be the beginning and will provide a starting point for others to expand these types of offerings for users of web archives.
\end{abstract}

\maketitle

\section{Introduction}

The web has become an integral part of our lives, shaping how we get news, shop, and communicate.  The web is also dynamic; webpages that existed two years ago may not exist today \cite{lim2001characterizing, klein-plosone2014}, webpages that exist today may not exist in two years, and even if webpages continue to exist, they can have significantly different content than before \cite{lexsig:ecdl08, jones-plosone2016}. Because of this, humanities scholars and social scientists are recognizing that web archives, such as the Internet Archive\footnote{\url{https://web.archive.org}}, are essential resources for their research \cite{starbird-2015, starbird-blog16, arora-15, milligan16}.

A goal of the Internet Archive is to save as much of the web as possible.  Other organizations, including libraries and museums, are concerned with creating focused collections of archived webpages about particular topics, such as human rights\footnote{\url{https://hrwa.cul.columbia.edu/}} and art history\footnote{\url{http://www.nyarc.org/content/web-archiving/}}. Humanities scholars may be interested in obtaining an overview of the topic of a particular webpage in a collection or in observing how that webpage has changed over time.  Many web archive interfaces provide only a textual list of archived versions, or \emph{mementos}, of a webpage, which requires that each one be explored individually.  This can be a tedious process as the number of mementos for each webpage grows.

As a first step to address this problem, we have developed visualizations that allow users to observe how a single webpage changes over time without the user accessing each memento individually. We do this by using the list of mementos of a webpage and choosing to display the screenshots, or thumbnails, of the most unique mementos, as a visual summarization. In addition to aiding researchers, this would also be useful in educating the general public about the temporal and dynamic nature of the web.  

\section{Viewing Mementos of a Webpage Over Time}

We use Memento \cite{memento-rfc} terminology, where URI-R identifies a web resource (or, original webpage), URI-M identifies that web resource at a particular point in time (or, a memento), and URI-T identifies its \emph{TimeMap}, which is a list of all of the mementos (URI-Ms) for the original resource (URI-R). Having a visual representation, or thumbnail, of each of the mementos in a TimeMap is helpful in quickly determining both the overall focus of the webpage and also how the design of that webpage has changed over time.  Other work on displaying webpage change has also included thumbnail views of individual mementos \cite{padia-jcdl12, adar-uist08,glam-workbench}.  But a problem arises when there are too many thumbnails to display.  For example, the Internet Archive has over 40,000 mementos for \url{bbc.co.uk} over a 20-year span. So even if the Internet Archive has thumbnails for all the mementos, some form of sampling will be necessary because the cognitive load of processing all the mementos will be beyond what the user can handle \cite{Mayer2003}. In addition, for webpages that do not change frequently, there may be many mementos with the same content, resulting in duplicate thumbnails. 

\begin{figure*}[ht!]
\centering
\subcaptionbox{700 thumbnails\label{fig:appl}}
{\includegraphics[width=0.45\textwidth]{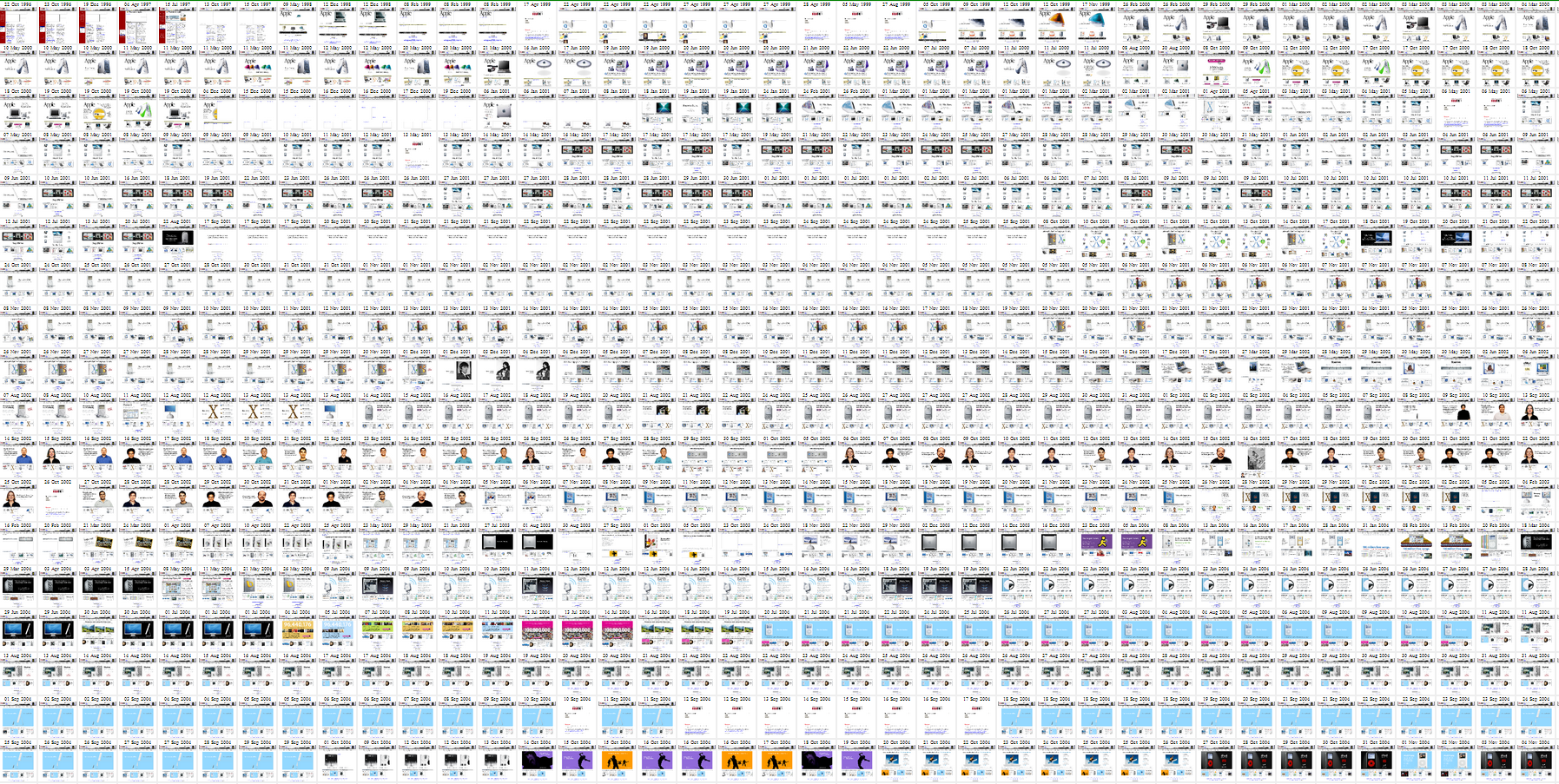}}
\subcaptionbox{69 thumbnails, sampled by our algorithm\label{fig:appl_threshold}}
{\includegraphics[width=0.3\textwidth]{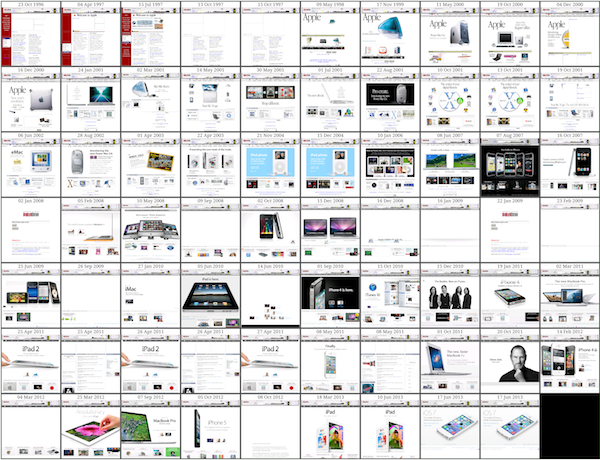}}
\caption{Thumbnails from the www.apple.com TimeMap}
\label{fig:apple-combined}
\end{figure*}

In our previous work \cite{alsum-ecir14}, we developed a technique to determine which mementos are useful to display. The algorithm compares the amount of change in the HTML  of consecutive mementos and selects the most unique. This process is much more efficient, in both time and space, than generating all thumbnails and then performing image processing. In this work, we implement this technique for TimeMap visualization.  For example, in Figure \ref{fig:apple-combined} we show thumbnails of 700 mementos of \url{www.apple.com} (we were unable to capture a figure showing all 8000 mementos) and a set of 69 thumbnails chosen by our algorithm from the full set of 8000. There are many duplicate thumbnails in Figure \ref{fig:appl} that have been eliminated in Figure \ref{fig:appl_threshold}.

\section{Visualizations}

We have implemented four different visualizations to show how a webpage changes over time: image grid, image slider, timeline, and animated GIF. The initial implementations of the visualizations were developed by Shankar \cite{shankar-ms17}.  In addition to sampling over the full TimeMap, we allow users to select a time range over which the summarization should be generated. In all of these visualizations, the thumbnail summarization algorithm will choose the thumbnails to display.  The visualizations are just different ways to present and interact with the same information.

\subsection{Image Grid}
\begin{figure*}[ht]
\centering
\includegraphics[width=0.75\textwidth, frame]{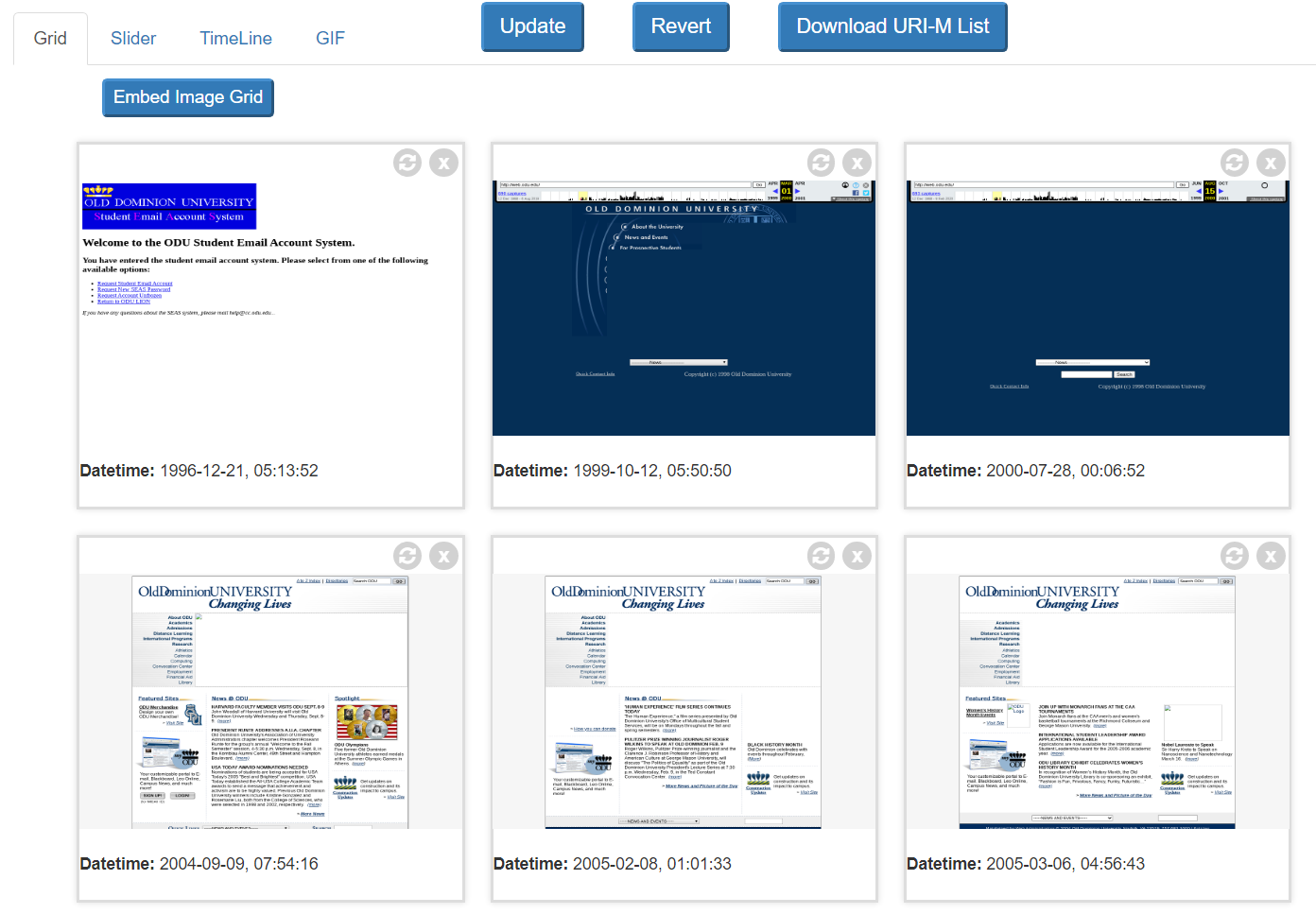}
\caption{Image Grid for http://www.odu.edu/}
\label{fig:image-grid}
\end{figure*}

The Image Grid shows the thumbnails of all the unique mementos arranged in a left to right, top to bottom manner. Clicking on the thumbnail loads the source memento, giving the user the opportunity to explore further. On the top right of each thumbnail in the grid is a refresh button to allow users to regenerate the thumbnail if it appears incomplete and an `X' to allow users to remove thumbnails from the visualizations. Figure \ref{fig:image-grid} shows an example of the grid view for \url{http://www.odu.edu/}.

\subsection{Image Slider}
\begin{figure*}[ht]
\centering
\includegraphics[width=0.55\textwidth, frame]{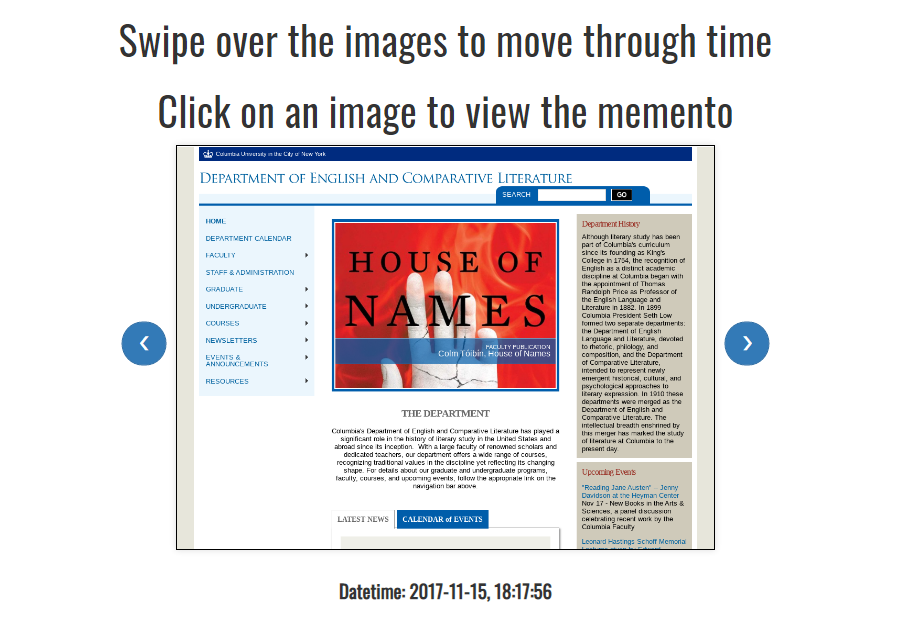}
\caption{Image Slider for http://columbia.edu/cu/english/}
\label{fig:image-slider}
\end{figure*}

The Image Slider imitates the photo slider functionality used in Apple's iPhoto\footnote{\url{http://web.archive.org/web/20150101033528/http://apple.com/mac/iphoto/}}. By moving the cursor across the thumbnail image, the next thumbnail is displayed. As with the Image Grid, clicking on the thumbnail loads the source memento. The user can cycle through the thumbnails by clicking arrow buttons to the left and right of the slider. Figure \ref{fig:image-slider} shows a static example of the image slider for \url{http://columbia.edu/cu/english/}.

\subsection{Timeline}
\begin{figure*}[ht]
\centering
\includegraphics*[width=0.9\textwidth, frame]{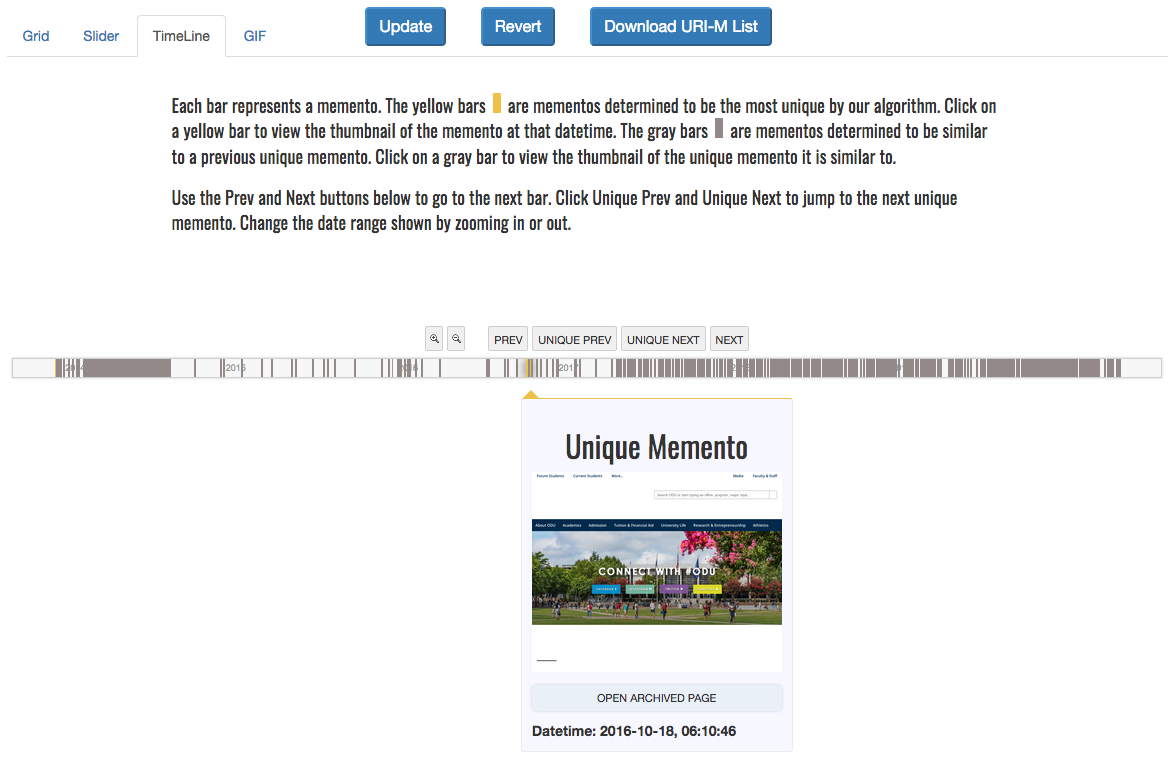}
\caption{Timeline view for http://www.odu.edu/}
\label{fig:timeline}
\end{figure*}

The Timeline view arranges the thumbnails according to the mementos' datetimes (time of capture). The Timeline view is equipped with zoom, next, previous, next unique, previous unique buttons to allow the user to easily navigate between the unique and regular mementos. The unique mementos are represented with yellow stripes and the regular ones are represented by gray stripes on the timeline. Non-unique mementos do not have a screenshot, but are represented by the thumbnail of the previous unique memento with an indication that this memento is ``similar to'' that memento.  The Timeline view is based on Timeline Setter library\footnote{\url{http://propublica.github.io/timeline-setter/}} \cite{timeline-propublica}, developed by ProPublica. Figure \ref{fig:timeline} shows an example of the Timeline view for \url{http://www.odu.edu}.

\subsection{Animated GIF}
\begin{figure*}[ht]
\centering
\includegraphics[width=0.6\textwidth, frame]{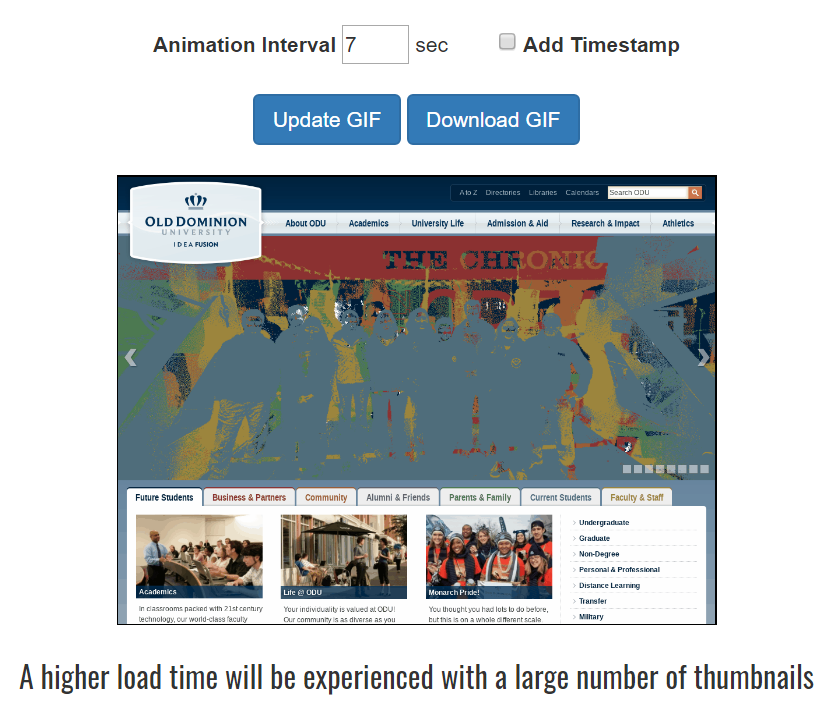}
\caption{Animated GIF for http://www.odu.edu/}
\label{fig:gif}
\end{figure*}

The Animated GIF visualization, shown in Figure \ref{fig:gif}, takes the thumbnails from the Image Grid and uses the GifShot library \cite{gifshot} to create an animated GIF. This GIF can be downloaded by clicking the ``Download GIF'' button. The user is given the option to include a timestamp on each screenshot in the GIF. When selected, a watermark of the appropriate datetime is added to each screenshot. This visualization also allows the user to adjust the time interval in seconds between each frame of the animated GIF. If any settings are changed, the GIF will be updated once the user presses the ``Update GIF'' button.

\section{Selecting Unique Thumbnails}
We select the most unique thumbnails based on a comparison of the SimHash \cite{simhash-2002} of each memento's HTML source. To improve the response time for very large TimeMaps (over 1000 mementos), we first sample a maximum of 1000 mementos before computing the SimHash and continuing the selection algorithm. These techniques are described further below. The initial implementation of the base summarization algorithm was developed by Kelly \cite{thumbnail-code}.

\subsection{SimHash}
AlSum summarization \cite{alsum-ecir14} relies on comparing the HTML source of mementos rather than images of the rendered webpage. Downloading and analyzing just the HTML source is much more efficient in both time and storage than rendering the entire webpage, taking a screenshot, and comparing images. Because the rendering process is expensive, we only want to render and take screenshots of those mementos we deem to be most unique.  To do this, we download the HTML content of all mementos in the TimeMap and compute the SimHash of each memento. Then we compare the SimHashes of two mementos at a time, computing the Hamming distance \cite{hamming:1950}
between the two SimHashes. We can use different Hamming distance thresholds to require different amounts of changes and thus produce different numbers of representative thumbnails.

\begin{figure*}[ht]
\centering
\includegraphics[width=0.8\textwidth, frame]{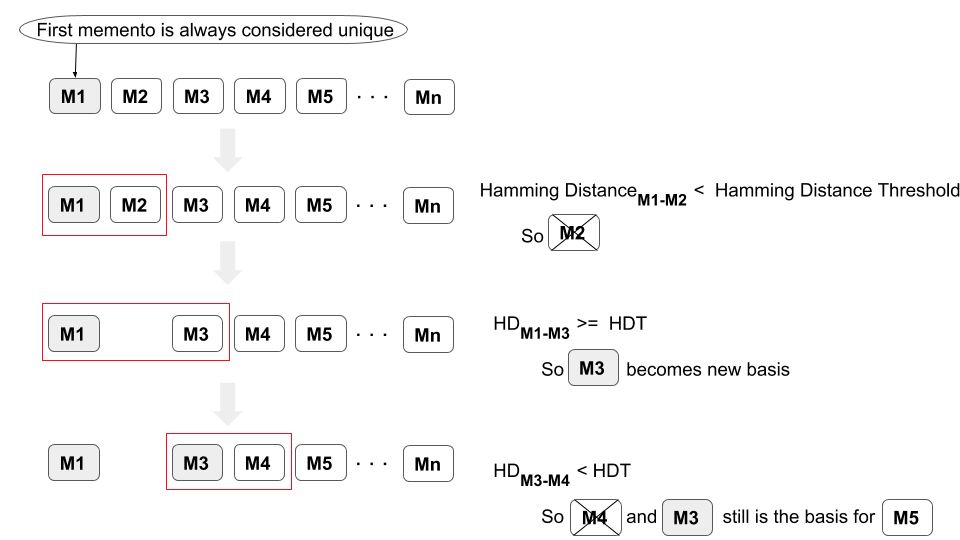}
\caption{Determining the most representative mementos using the Hamming distance}
\label{fig:hamming}
\end{figure*}

The process of choosing the best thumbnails using SimHash is outlined in Figure \ref{fig:hamming}. The filtering process begins by selecting the first memento in the TimeMap to be included in the summarization. This memento will act as a baseline to compare to subsequent mementos. In Figure \ref{fig:hamming}, the Hamming distance threshold is denoted as HDT, and each memento is M1, M2, ... Mn. The first memento, M1, is compared to each consecutive memento until the Hamming distance is greater than or equal to the threshold. In the diagram, this occurs when M1 is compared to M3.

\begin{figure*}
    \centering
    \begin{tabular}{ccc}
        \includegraphics[width=0.3\textwidth]{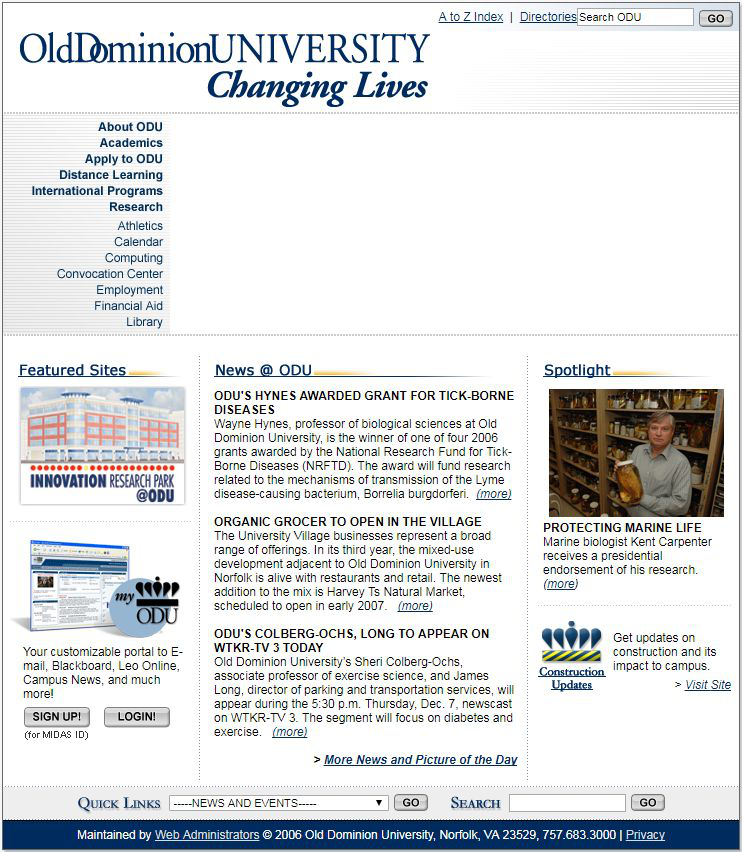} &   \includegraphics[width=0.3\textwidth]{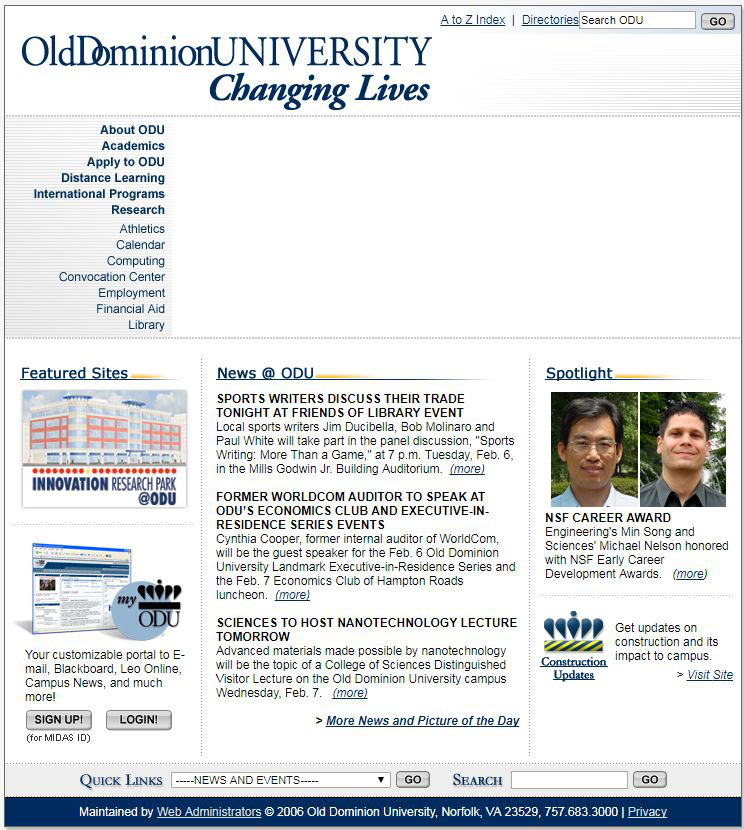} & 
        \includegraphics[width=0.3\textwidth]{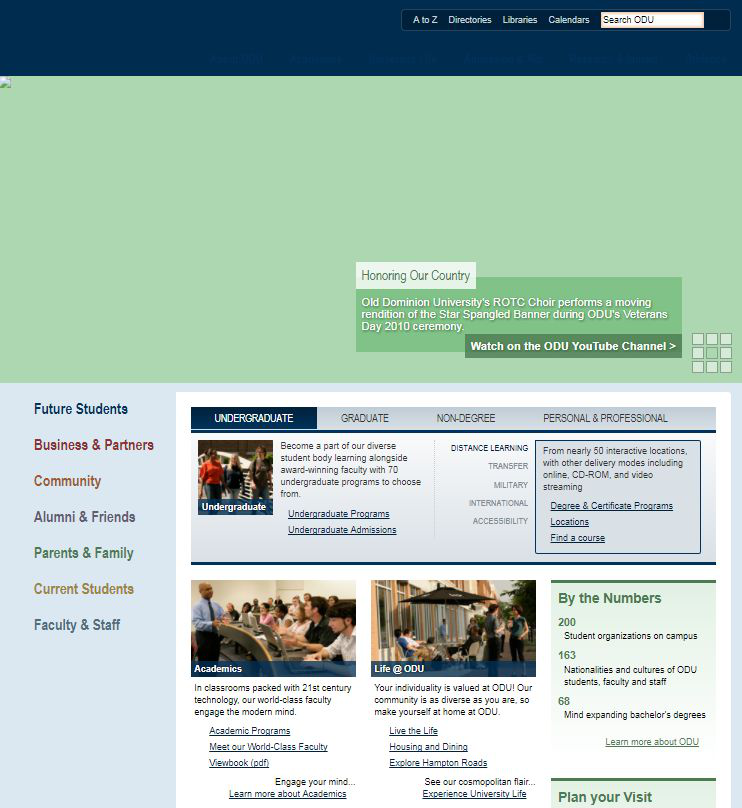} \\
        (a) M1 & (b) M2 & M3 \\
    \end{tabular}
    \caption{Thumbnails of M1, M2, M3 from the SimHash example}
    \label{fig:simhash}
\end{figure*}

To demonstrate why SimHash is a better measure in calculating the webpage similarities than other hash techniques like MD5, we consider three mementos, M1,  M2, and M3,  from the TimeMap of \url{http://www.odu.edu}. 
The thumbnails are shown in Figure \ref{fig:simhash}.
The thumbnails for mementos M1 and M2 look similar, but M3 is distinct from the other two. Upon closer look, M1 and M2 contain different content, though the layout is the same. First, we compute the MD5 hashes of the HTML content of these mementos: 
\newline
\texttt{
\newline
\$ curl URI-M1 | md5sum -> fc8e53aebb9061f390aba82665581295
\newline
\$ curl URI-M2 | md5sum -> d546e192eab633f4d1b4451399c8adcc
\newline
\$ curl URI-M3 | md5sum -> 5e98bc5367c86f3ffaea0b8c3deb3f5d
\newline
}
\newline
After the hashes are generated, the Hamming distance is calculated between the pairs (M1, M2) and (M2, M3). The Hamming distance is the minimum number of substitutions required to convert one string to another. The results are as shown below:
\newline
\texttt{
\newline
\$ node hammingdistance M1 M2 -> 30
\newline
\$ node hammingdistance M2 M3 -> 32
\newline
}
\newline
For MD5, we see that the distances between (M1, M2) and (M2, M3) are similar, meaning that MD5 considers them to have similar amounts of difference.

Now we repeat the same process using SimHash as the hashing technique: 
\newline
\texttt{
\newline
\$ curl URI-M1 | SimHash -> 8c27981eaed151cfa645ad823932eac6
\newline
\$ curl URI-M2 | SimHash -> 8c27981\textcolor{red}{fa}a\textcolor{red}{d}951cf\textcolor{red}{8}645ad823\textcolor{red}{d}32eac\textcolor{red}{2}
\newline
\$ curl URI-M3 | SimHash -> fa3799170258494b9443b9be3977a84e
\newline
}
\newline
The Hamming distances are calculated as below:
\newline
\texttt{
\newline
\$ node hammingdistance M1 M2 -> 6
\newline
\$ node hammingdistance M2 M3 -> 27
\newline
}
Here we see that with SimHash, the distance between (M1, M2) is much smaller than the distance between (M2, M3). This reflects the notion that we have after observing the screenshots of the rendered pages. Hence we can conclude that SimHash correlates better with the webpage similarity.

\subsection{TimeMap Sampling Algorithm} 
\begin{figure*}[ht]
\centering
\includegraphics[width=0.9\textwidth, frame]{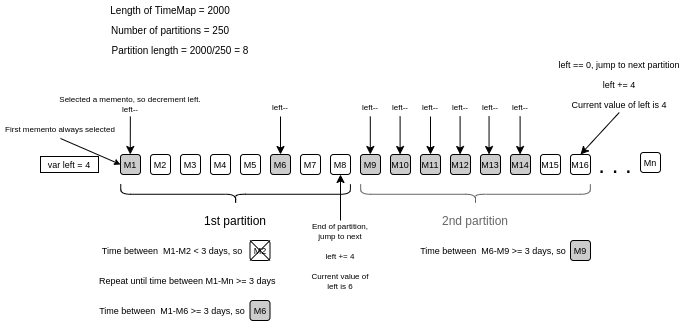}
\caption{TimeMap Sampling Algorithm}
\label{fig:sampling-alg}
\end{figure*}

Some TimeMaps contain hundreds of thousands of mementos. To efficiently calculate the most representative mementos from such large TimeMaps, we take a sample of up to 1000 mementos from the TimeMap to analyze. The algorithm used to extract the sample of mementos is illustrated in Figure \ref{fig:sampling-alg}. 
The TimeMap is split into 250 equal partitions. For example, if the TimeMap contains 2000 mementos, each partition will contain 8 mementos. From each partition, up to four mementos will be chosen. The algorithm always chooses the first memento in the set. Each time a memento is chosen, a counter of the number of mementos left that can be selected from the partition is decremented. The datetime of the first memento is then compared to the datetime of the second memento. If the time between the mementos is less than 3 days, the second memento will be skipped. The first memento is compared to each consecutive memento until the distance between datetimes is at least 3 days. As shown in Figure \ref{fig:sampling-alg}, this occurs when the first memento is compared to the sixth memento. If the end of the partition is reached before the counter is 0, then four mementos plus the leftover mementos can be chosen from the next partition. In Figure \ref{fig:sampling-alg}, only two mementos were selected from the first partition, so six mementos can be selected from the second partition. This process continues until the end of the TimeMap is reached. 

\begin{figure*}
\centering
\includegraphics[width=0.9\textwidth]{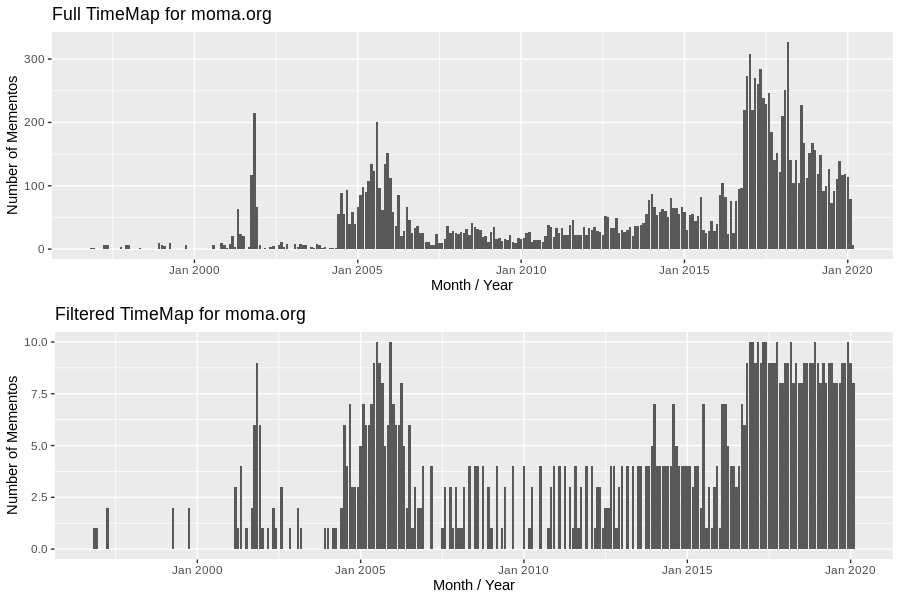}
\caption{Sample selected from http://moma.org/}
\label{fig:moma-sample}
\end{figure*}

The algorithm is designed to extract a sample that represents the distribution of the original TimeMap. For example, if most of the mementos in a particular TimeMap are from 2016, then most of the mementos in the sample will also be from 2016. Figure \ref{fig:moma-sample} shows a histogram of the TimeMap for moma.org compared to a histogram of the filtered TimeMap. The histograms are binned by month and year of the memento datetime.
It can be seen from the histograms that the filtered TimeMap has proportions similar to the original TimeMap. Sections of the TimeMap where hundreds of mementos are from the same month are omitted. This allows the unique mementos to be calculated more quickly since hundreds of similar mementos do not have to be compared.

\section{Caching TimeMaps and Thumbnails}

The determination of which thumbnails to choose as the most representative for a particular TimeMap may change over time, as new mementos are archived. For efficiency, we do not want to recompute SimHash values for mementos we have already processed, nor do we want to re-render thumbnails when those have already been generated. Thus, we have developed a process for caching TimeMaps, SimHash values, and thumbnail images. 

\subsection{TimeMap Caching} 
\begin{figure*}[ht]
\centering
\includegraphics[width=0.9\textwidth]{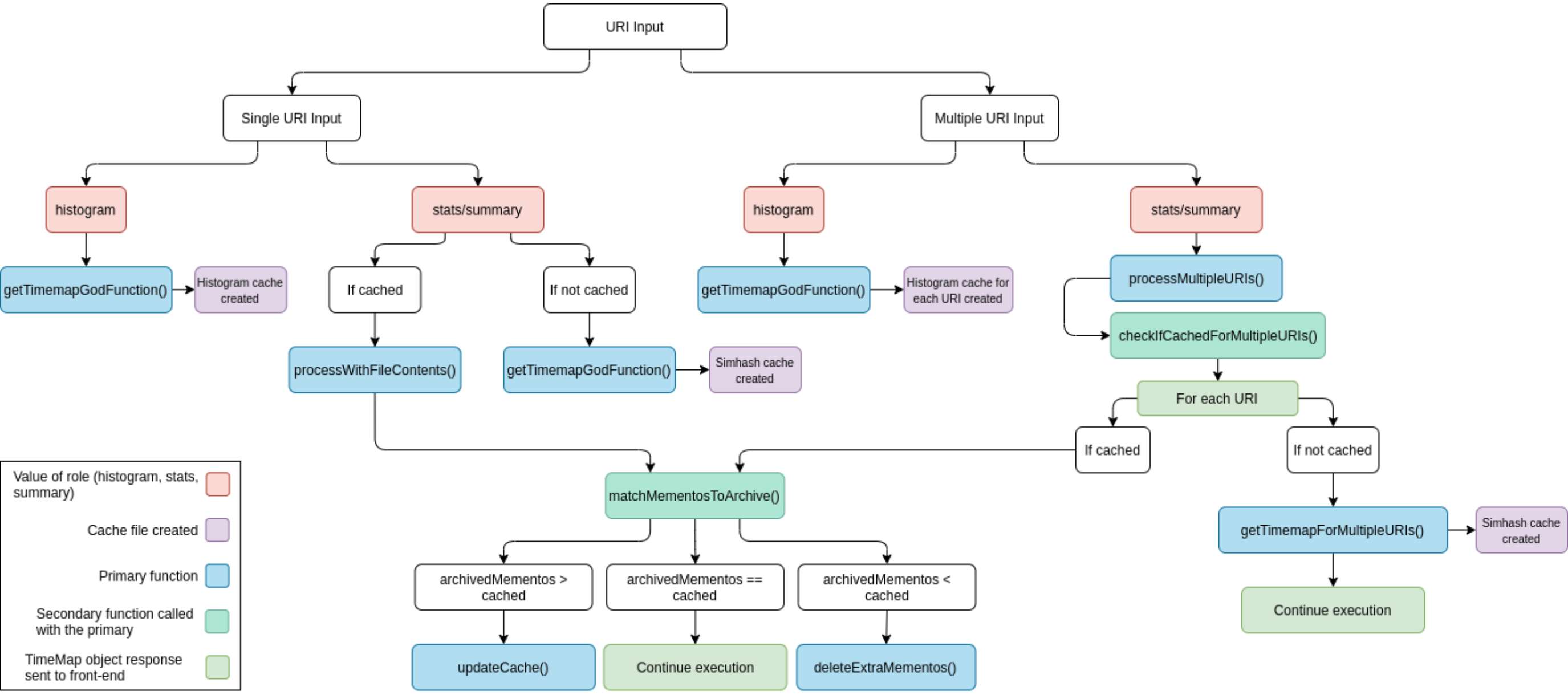}
\caption{Cache File System for TMVis}
\label{fig:cache-system}
\end{figure*}

As the web archiving process continues, a TimeMap for a webpage is likely to continue to grow. In addition, sometimes archives might even delete a memento from a TimeMap either temporarily or permanently. We have accounted for these two instances by checking if the number of mementos we have cached for a particular URI-R matches the number of mementos the archive has for that URI-R. In order to ensure that the user is shown the most up to date information, we have developed the caching system shown in Figure  \ref{fig:cache-system} that stores current thumbnail summary information and updates the cache files as needed. 

When the user inputs a URI-R for processing, the full TimeMap is fetched and stored in a cache file. This cache file stores an array of datetimes for each memento in the TimeMap. The naming convention of this file is \texttt{histogram\_[archive name][collection identifier][URI-R]}. The user will then select the date range of mementos that they would like to view. After selecting a date range, the system checks if a SimHash cache file exists for the URI-R. If the file does not exist, the TimeMap is fetched and a new TimeMap object containing all of the mementos in the TimeMap is created. The mementos in the TimeMap object are then filtered down to consist of only mementos in the user-requested date range. If there are more than 1000 mementos after filtering down to the date range, the mementos are filtered again with the TimeMap sampling algorithm (described earlier) so that a sample of up to 1000 mementos remains. The SimHashes for the mementos in this sample will then be calculated and cached.

If the cache file does already exist for the specified URI-R, the file contents are read into a new TimeMap object. Since the mementos are from a previously fetched TimeMap, the cached mementos must be compared to the current TimeMap. To do this, the mementos in the histogram cache file are read into an array. This array is filtered, by both date range and with the TimeMap sampling algorithm, along with the mementos in the TimeMap object created from the previous file contents. If the TimeMap has remained unchanged since the previous SimHash cache file was created, then the mementos in the histogram cache array and the TimeMap object will be the same and execution may continue. If the histogram array contains more mementos than the TimeMap object, then the ``Update Cache'' process begins. Conversely, if the histogram array contains fewer mementos than the TimeMap object, then some mementos have been removed from the archive. These mementos must also be removed from the TimeMap object as the system may not be able to take a screenshot of a memento that no longer exists. Here the ``Delete Extra Mementos'' process begins (as described below). The purpose of this is to reduce the need for recomputing the SimHashes. Iterating through the memento lists can be completed in a matter of seconds. However, recomputing the SimHashes can take several minutes. 

\subsection{Update Cache Process}
The ``Update Cache''  process updates the existing SimHash cache file for a given URI-R. This process starts by fetching the URI-R's TimeMap from the archive. A TimeMap object reads this data and filters mementos for date range if needed. If the number of mementos is greater than 1000, they are filtered down to include only a sample of 1000. The mementos that exist in the cache file are deleted from the object and stored in another object for later use. The SimHashes of all the mementos not in the cache file are calculated and then merged into the object containing the old cached mementos. This object is then sorted old to new by date and the cache file is overwritten with these mementos.

\subsection{Delete Extra Mementos}
The ``Delete Extra Mementos'' process is triggered when the number of cached mementos exceeds the number of mementos currently in the TimeMap. The datetimes of the mementos in the cached mementos array are compared to those in the archived mementos array. If a datetime from the cached array does not match a datetime in the archived array, the memento in the cached array is considered to be currently removed from the archive and therefore discarded from the array. After removing the cached memento from the array, the loop continues comparing the datetimes. The discarded mementos are not deleted from the cache file since they may be added back to the archive in the future. By the end of this process, the cached memento list and the archived memento list will be identical.

\section{System Walkthrough}

We have developed a web service that allows general users to view an overview of any TimeMap with the developed visualizations.  The initial implementation of the web service was developed by Gunnam \cite{gunnam-ms18}. We have also developed embeddable versions (with simple embed scripts like those provided by Twitter, YouTube, etc.) that allows webpage authors to include these visualizations in their own webpages.  The web service is available at \url{http://tmvis.cs.odu.edu/}, the source code is hosted on GitHub at \url{https://github.com/oduwsdl/tmvis}.

\subsection{TimeMap Selection}
\begin{figure*}[ht]
\centering
\includegraphics[width=0.75\textwidth]{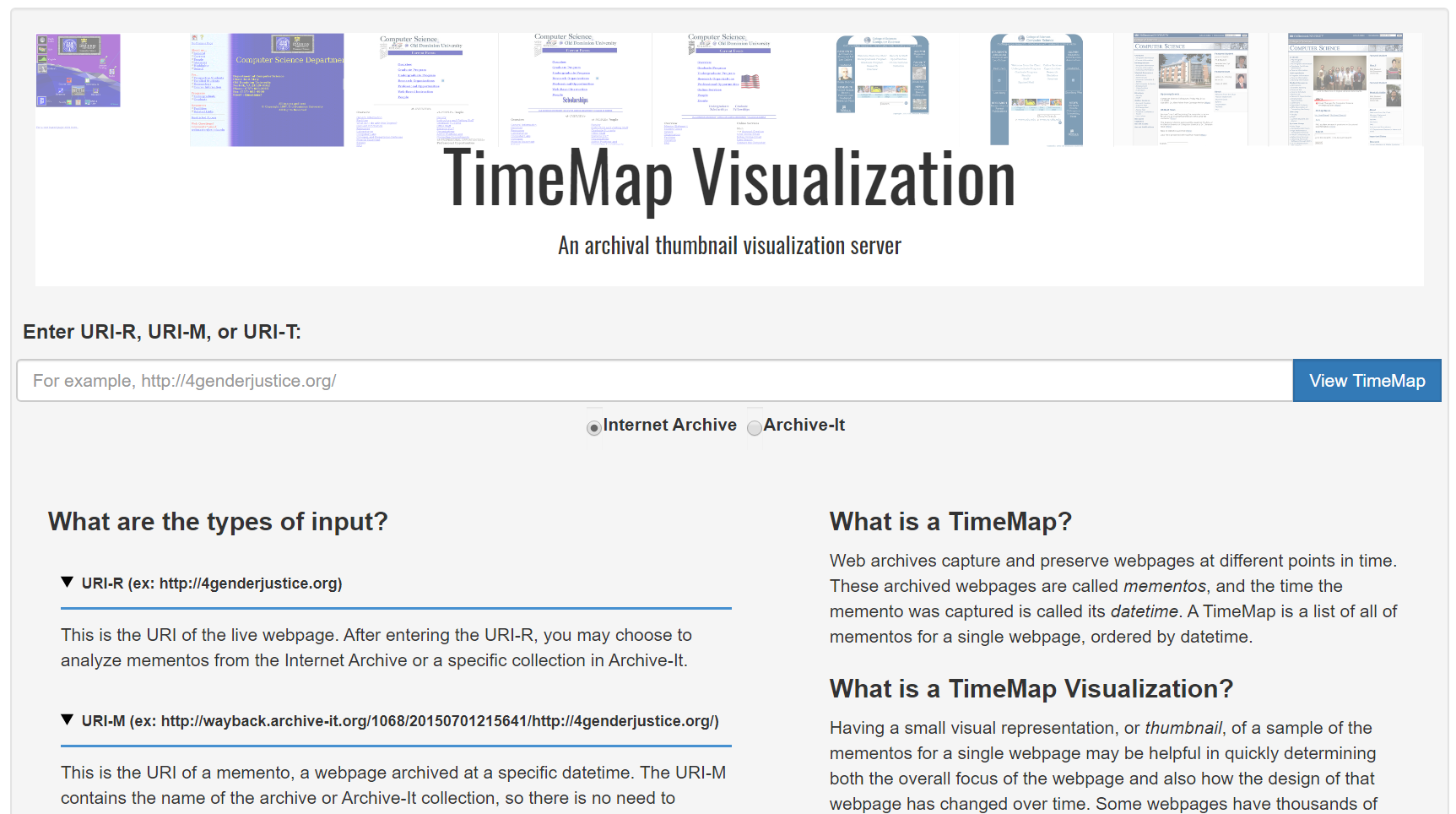}
\caption{Home page of the service http://tmvis.cs.odu.edu/}
\label{fig:main-page}
\end{figure*}

From a user's perspective, there are three phases needed to summarize the webpage. Figure \ref{fig:main-page} shows the first phase, the home page of the application where the user can input the URI-R, URI-M, or the URI-T. Again, according to Memento \cite{memento-rfc} terminology, URI-R identifies an  original webpage, URI-M identifies a memento, and URI-T identifies a TimeMap. The user can choose the web archive to be used, either the Internet Archive  or Archive-It, a collection and subscription-based archival service operated by the Internet Archive. Opting for Archive-It allows the user to include a collection number as an additional parameter. If no collection number is input, the value ``all'' is considered to be the collection number.  We currently only support the Internet Archive and Archive-It, but the list of archives could be expanded to any Memento-compatible public web archive or Memento aggregator \cite{alam2016memgator, timetravel}. 

\begin{figure*}
\centering
\includegraphics[width=0.75\textwidth]{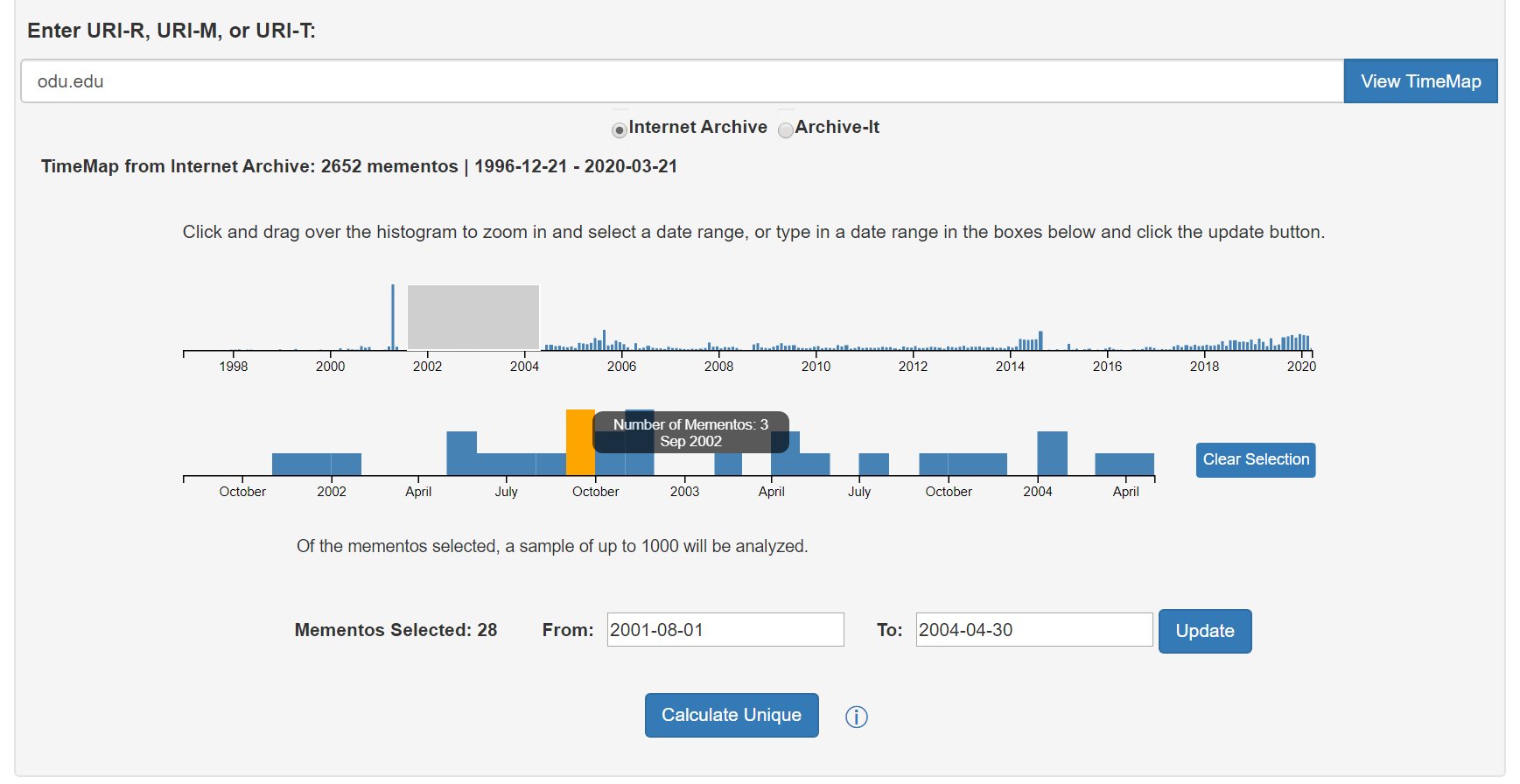}
\caption{Histogram page of the service http://tmvis.cs.odu.edu}
\label{fig:histogram-page}
\end{figure*}

The second phase in the summarization process begins with the user clicking the ``View TimeMap'' button after providing the proper input. The TimeMap is then requested from the appropriate archive and cached for future reference. The user is presented with a histogram displaying the number of mementos available over time as shown in Figure \ref{fig:histogram-page}. The user may use the histogram to select a date range in the TimeMap to summarize, or they may choose to summarize the entire TimeMap. To choose a date range, the user may either click and drag over the histogram to select the desired range, or they can type the date range into the boxes below. 

Some URI-Rs have very few mementos and so users may be interested in viewing all of these mementos. Since processing just a few mementos can be done rather quickly, for URI-Rs with less than 12 mementos, the user is given the option to process and display thumbnails for all mementos by clicking the ``Generate All Thumbnails'' button.  

\subsection{Comparing Mementos}
\begin{figure*}[ht]
\centering
\includegraphics[width=0.75\textwidth]{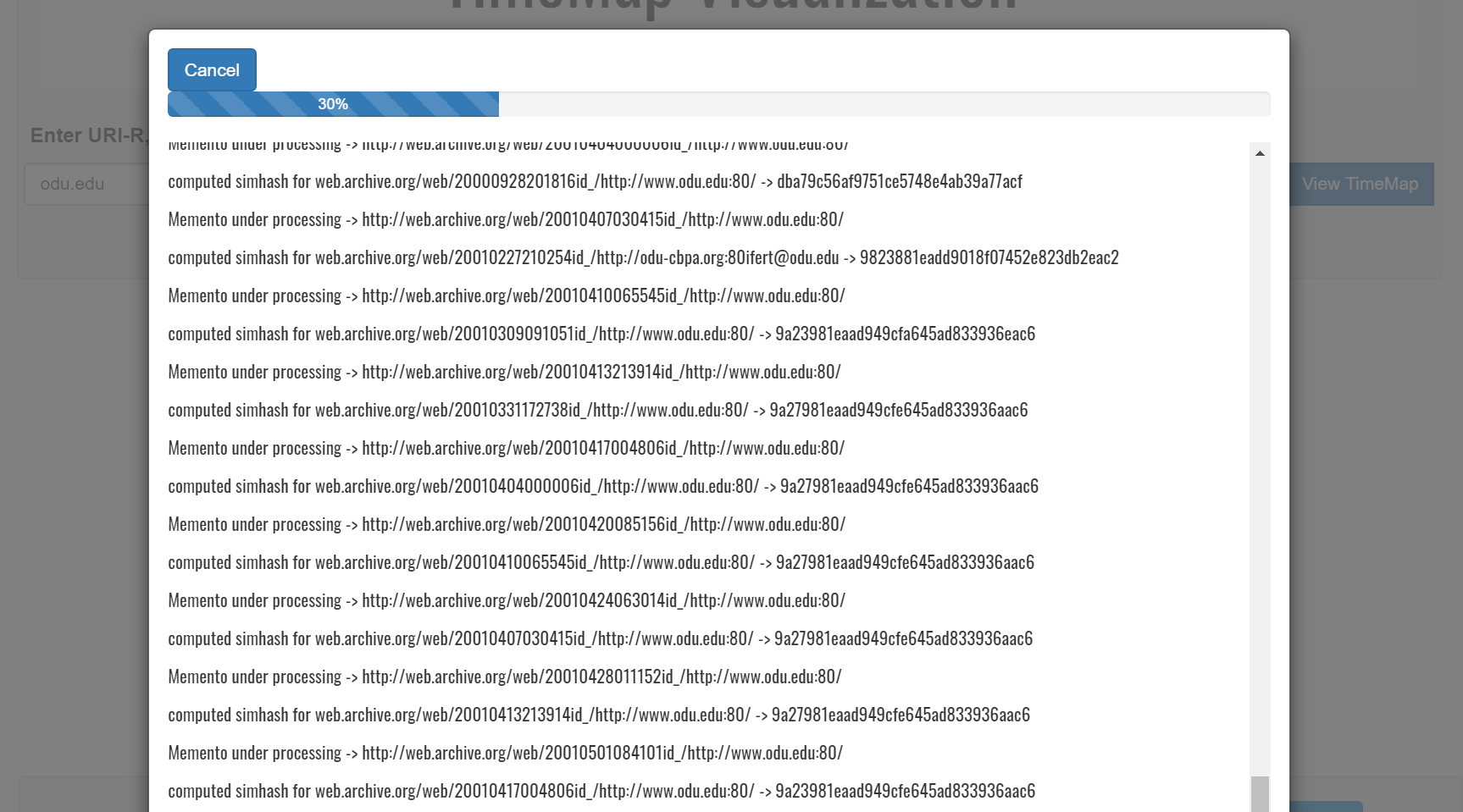}
\caption{Home page of http://tmvis.cs.odu.edu with progress window}
\label{fig:loading-window}
\end{figure*}

When the user clicks the ``Calculate Unique'' button, the web service downloads the HTML source and computes the SimHash of the requested mementos. Then the SimHashes of the mementos are compared according to AlSum et al. \cite{alsum-ecir14}. The user is then notified with the continuous stream of events taking place at the server through a progress window, as shown in Figure \ref{fig:loading-window}.

\begin{figure*}[ht]
\centering
\includegraphics[width=0.75\textwidth]{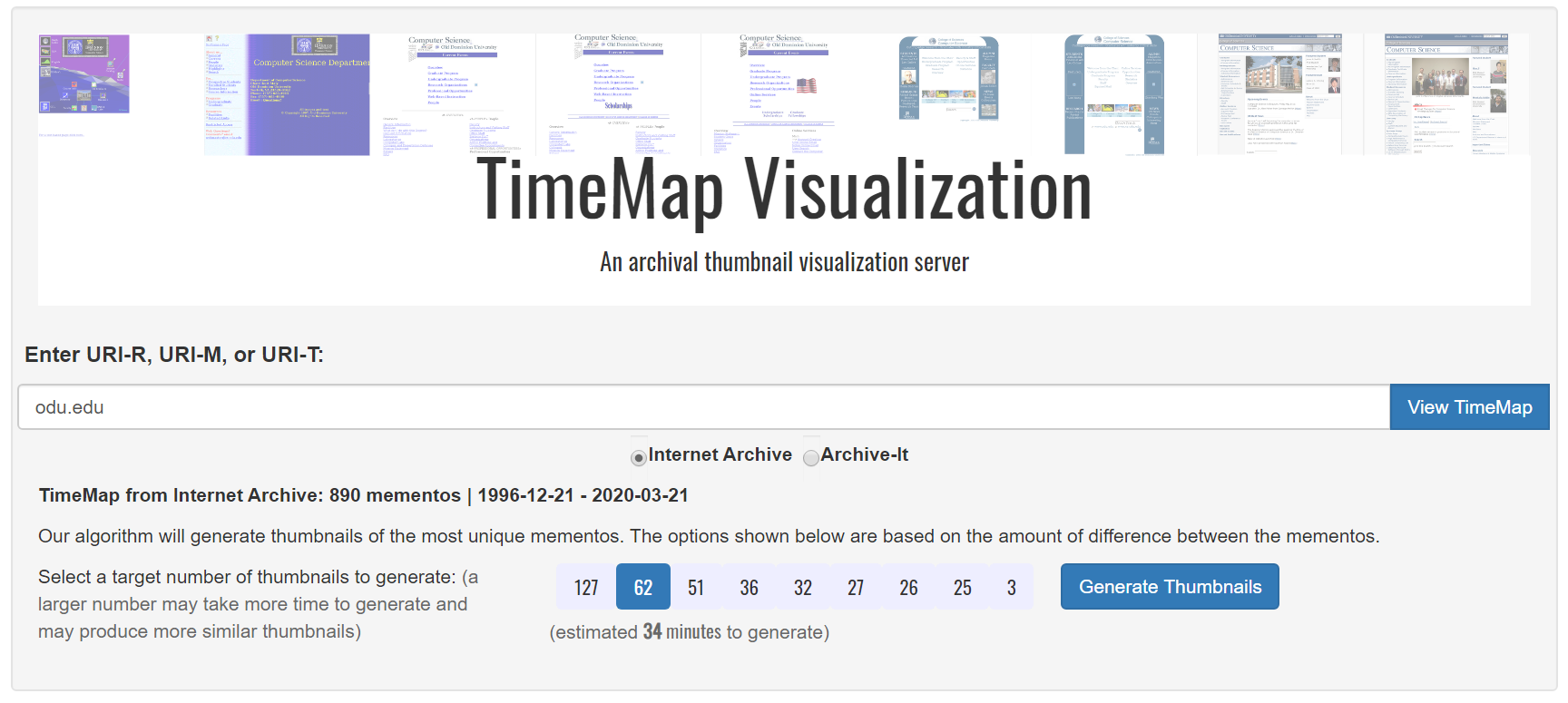}
\caption{tmvis.cs.odu.edu showing the number of unique thumbnails calculated for http://odu.edu/}
\label{fig:oduedu-stats}
\end{figure*}

Once the server side computation of unique number of representative thumbnails is complete, the user is presented with the date range of mementos under consideration along with options for choosing the number of representative thumbnails. Figure \ref{fig:oduedu-stats} shows the options available after processing  \url{http://odu.edu/} with the Internet Archive as the source. Figure \ref{fig:oduedu-stats} also shows that the TimeMap of \url{http://odu.edu} can be summarized using 127, 62, 51, 36, 32, 27, 26, 25, or 3 thumbnails. Choosing 127 thumbnails will summarize the TimeMap with more similar webpages and hence takes more time to generate. Choosing 3 thumbnails summarizes the TimeMap with more distinct webpages and takes a shorter amount of time to generate.

Once the representative memento filtering based on Hamming distance is complete, if the smallest Hamming distance produces a number of representative mementos greater than 9, then the option of three unique mementos will be given to the user in addition to the other options as shown in Figure \ref{fig:oduedu-stats}. These three mementos are the first, last, and central mementos from the smallest list of unique mementos that was obtained from the smallest Hamming distance threshold. If the list of unique mementos is even and there are two mementos in the center of the array, the central memento is considered to be the one with the lower index.

\subsection{Thumbnail Generation and Visualization}
The third and final phase of thumbnail generation starts when the user chooses the desired number of unique thumbnails and then clicks on the ``Generate Thumbnails'' button. The web service uses Puppeteer\footnote{\url{https://developers.google.com/web/tools/puppeteer}}, a headless web browser, to render each memento and capture a thumbnail screenshot. Once the thumbnails are captured for all the mementos, they are presented to the user with the four visualization widgets shown in Figures \ref{fig:image-grid}-\ref{fig:gif}. The default tab shows the Image Grid, however users can switch between tabs to view the other visualizations: Image Slider, Timeline, and Animated GIF.

\subsubsection{User Control Over Representative Thumbnail Selection}
\begin{figure*}[ht]
\centering
\includegraphics[width=0.9\textwidth]{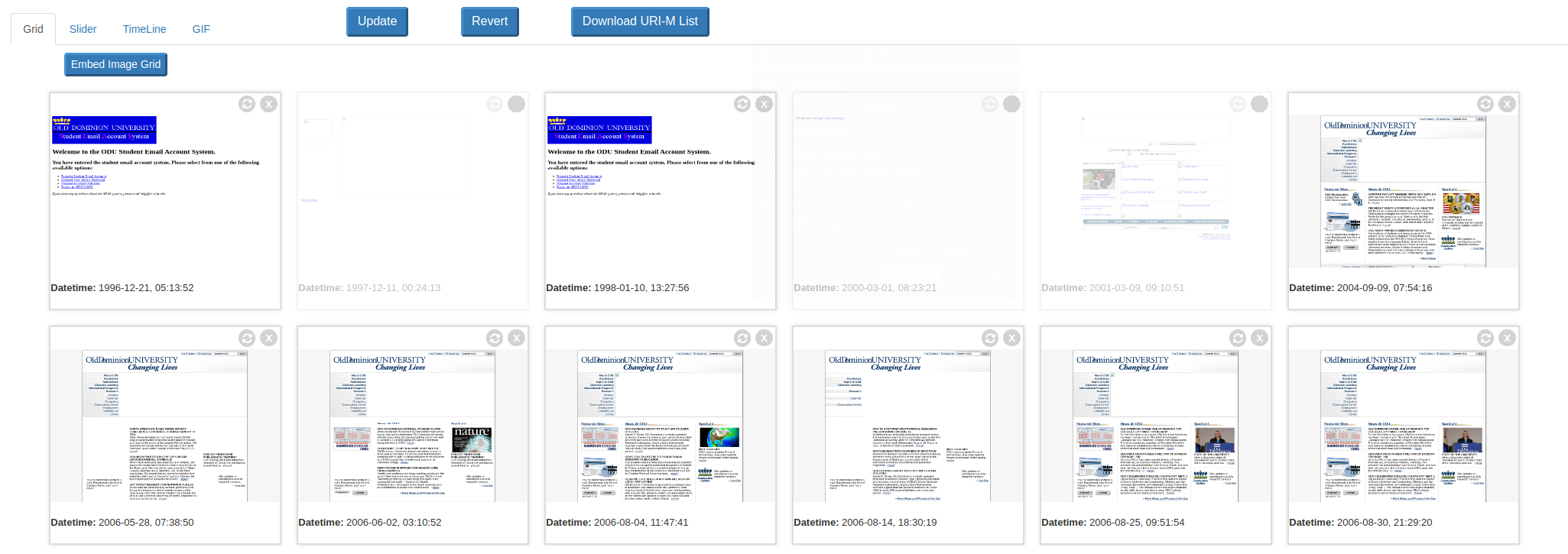}
\caption{Demonstrating thumbnails selected for removal}
\label{fig:image-grid-removal}
\end{figure*}

The Image Grid visualization tab allows the user to customize the thumbnails used for all of the visualizations. The user may remove thumbnails and refresh thumbnails. To remove a thumbnail, the user clicks the `X' button on the top right corner of each thumbnail they would like removed. When clicked, the thumbnail becomes grayed out, showing that it is marked for removal. The user may click the button again to deselect the thumbnail. Once the user has selected which thumbnails they would like excluded from the grid view, the user can click the ``Update'' button at the top of the Image Grid to apply the changes. If the user would like to add back the thumbnails, the user can click the ``Revert'' button at the top of the Image Grid to revert back to the original set of thumbnails. Figure \ref{fig:image-grid-removal} shows an example of the Image Grid for \url{http://odu.edu/} with three thumbnails selected for removal and the ``Revert'' and ``Update'' buttons at the top of the grid.

Mementos are loaded from web archives. The time to render a full webpage varies due to the number and types of embedded resources in the webpage and current load on the web archive server. Some mementos may take a long time to fully render, making it possible for the screenshot to be taken before all visual content appears on the page. This issue is accounted for by providing the user with the capability to request a new screenshot to be taken of a particular memento. This feature is incorporated into the user interface by placing a refresh icon in the top right corner of each memento in the Image Grid visualization, as shown in Figure \ref{fig:image-grid-removal}. When a request is made for a screenshot to be retaken, the system waits for a longer time period than the prior attempt to allow the visual content to appear on the page. A new thumbnail is then generated from this new screenshot and each visualization is updated to include the new thumbnail.

\subsubsection{Download URI-M List}
Once the list of representative mementos has been determined, this list is written to a text file. The user can download this file by clicking on the ``Download URI-M List'' button on the Image Grid tab. This text file may be used for many purposes, including as input to Raintale \cite{raintale}, a tool that allows users to create and share stories using archived webpages.

\subsubsection{Downloadable Animated GIF}
The Animated GIF visualization, which has already been described (Figure \ref{fig:gif}), is easily embeddable into other webpages. We have provided a ``Download GIF'' button that allows the user to download the GIF image file. The user can customize the animated GIF in several ways. The user is given the option to include a timestamp on each screenshot in the GIF. When selected, a watermark of the appropriate datetime is added to each screenshot and then the new Animated GIF is generated with the watermarked images. This visualization also allows the user to adjust the time interval in seconds between each frame of the animated GIF. If any settings are changed, the GIF will be updated once the user presses the ``Update GIF'' button.

\subsubsection{Embed Feature for Image Grid and Slider}
\begin{figure*}[ht]
\centering
\includegraphics[width=0.6\textwidth]{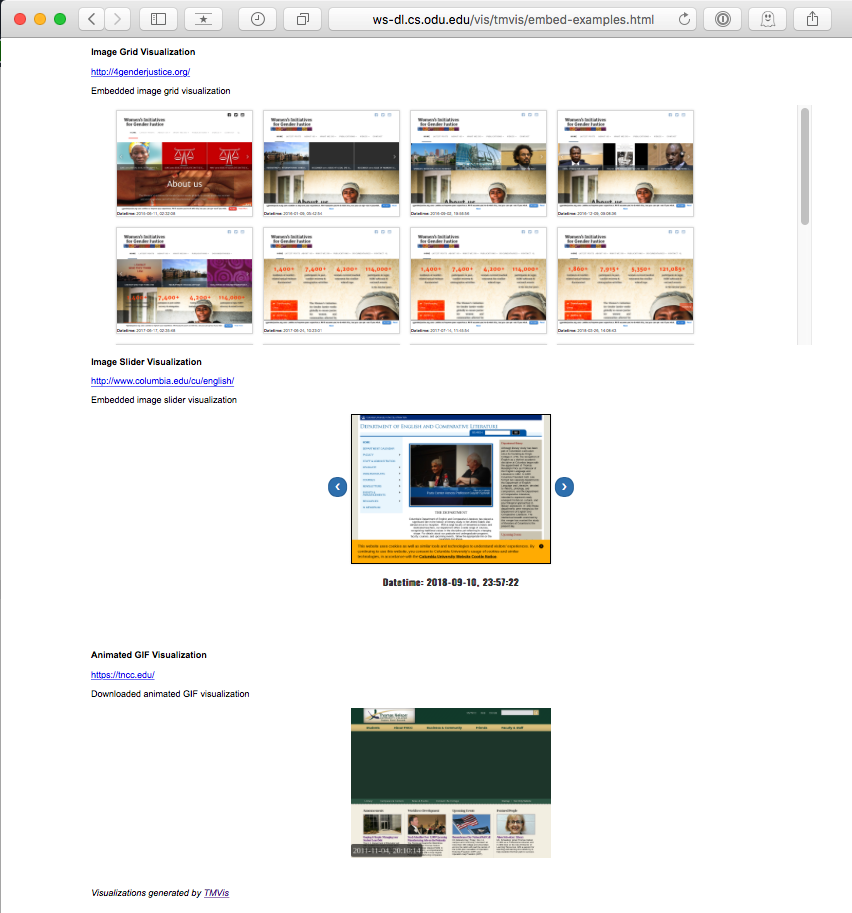}
\caption{Visualizations embedded in another webpage}
\label{fig:embed-example}
\end{figure*}

The Image Grid and Image Slider visualizations generated by TMVis give the user the option to embed the visualizations into other webpages. Users can click the ``Embed Image Grid'' and ``Embed Image Slider'' buttons to generate the embed code. Generating the embed code takes into account the user control over representative thumbnail selection by only including thumbnails that are currently being shown. This allows users to control which thumbnails they are sharing when they embed the visualization. The preferred visualization is encoded in an iframe with the selected thumbnails. For the Image Grid embed content, the `X' and refresh are removed.  An example of embedding the Image Grid, Image Slider, and Animated GIF in a simple webpage is provided at \url{https://ws-dl.cs.odu.edu/vis/tmvis/embed-examples.html} and shown in Figure \ref{fig:embed-example}.

\subsection{Multiple URI Input}
\begin{figure*}[ht]
\centering
\includegraphics[width=0.6\textwidth, trim=0 25 40 30, clip]{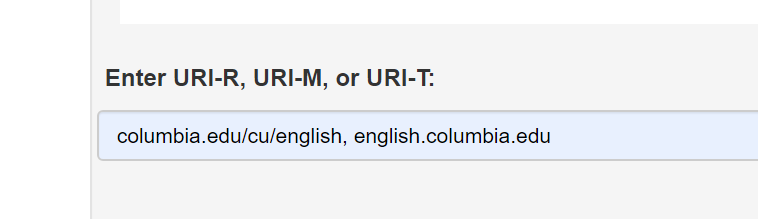}
\caption{Multiple URI-R input}
\label{fig:multiple-uri-input}
\end{figure*}

Some websites update their hostname, creating multiple URI-Rs that point to the same effective webpage. When this happens, any new mementos that are created after the hostname change may not correlate to the old hostname in the archive. Thus, there will be two separate TimeMaps. To accommodate this, users may input comma-separated URI-Rs. The system will fetch the TimeMap for each URI-R and merge them together, creating one TimeMap to process. This allows users to input all hostnames for a website and view all mementos associated with that particular website over time. For example, the user can input \url{http://columbia.edu/cu/english/}, \url{http://english.columbia.edu/} as shown in Figure \ref{fig:multiple-uri-input}.

When ``View TimeMap'' is clicked, the system checks for a cache file for either URI-R. If one is not cached, then the TimeMap for the URI-R is fetched. The SimHashes for that URI-R are calculated and cached for future use. If none of the URI-Rs are cached, this process is completed for each URI-R individually. For any URI-R that is cached, the cache file contents are read into the TimeMap object. Once all the mementos for each URI-R are merged, they are sorted by date.

\begin{figure*}[ht]
    \centering
    \begin{tabular}{cc}
        \includegraphics[width=90mm]{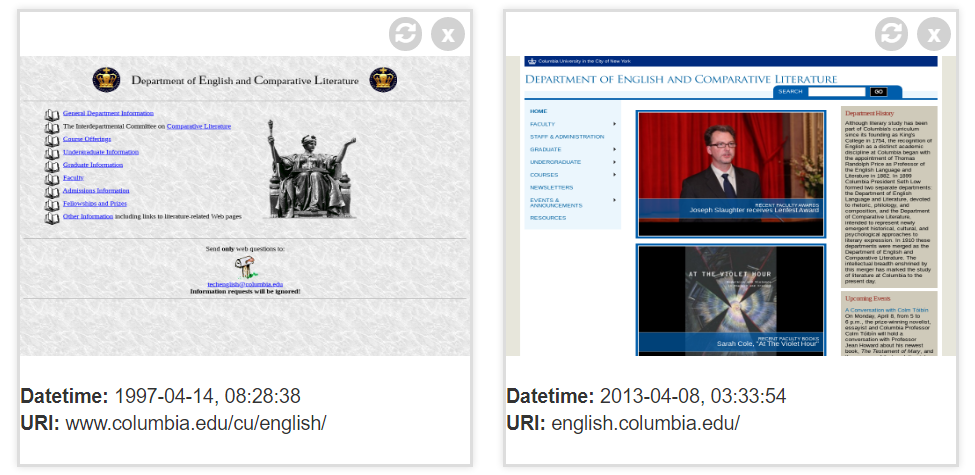} &   \includegraphics[width=60mm]{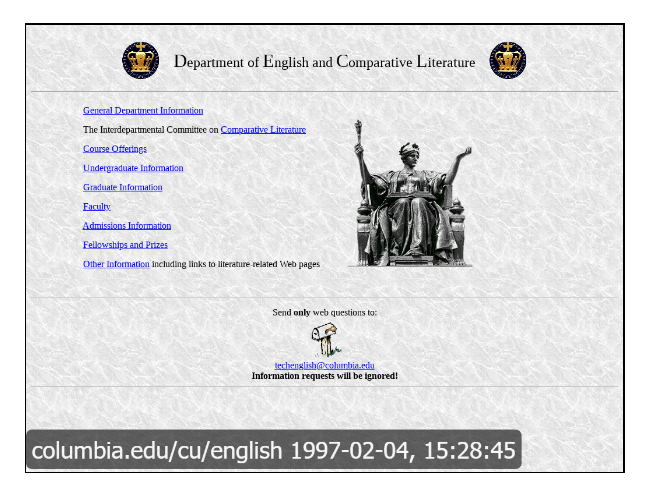} \\
        (a) URI-stamped Image Grid & (b) URI-stamped animated GIF \\
    \end{tabular}
    \caption{Visualizations with URI stamp functionality}
    \label{fig:uri-stamped-visualizations}
\end{figure*}

When the user loads the visualizations, the Image Grid and Animated GIF each have a URI stamp feature, as shown in Figure \ref{fig:uri-stamped-visualizations}. The URI stamp is so that the user knows exactly which URI-R each particular memento came from. On the animated GIF, the URI stamp is optional. It is not optional on the Image Grid.

\section{Conclusion}
We have presented a description of the implementation of TMVis, a web service for producing four visualizations of how a webpage changes through time. To do this we compare the difference in the SimHashes of the HTML source of pairs of mementos to determine which mementos are the most unique. Then we render and capture thumbnail-sized screenshots of the chosen mementos. These are then displayed as Image Grid, Image Slider, Timeline, and Animated GIF visualizations. The Image Grid, Image Slider, and Animated GIF visualizations can be embedded in other webpages. The web service is available at \url{http://tmvis.cs.odu.edu/}, and the source code is hosted on GitHub at \url{https://github.com/oduwsdl/tmvis}.

\section*{Acknowledgements}

This work has been supported by a NEH/IMLS Digital Humanities Advancement Grant (HAA-256368-17) \cite{weigle-neh-2017}. We are grateful for the support of the National Endowment for the Humanities (NEH) and the Institute of Museum and Library Services (IMLS), and for the input from our partners, Deborah Kempe from the Frick Art Reference Library and New York Art Resources Consortium and Pamela Graham and Alex Thurman from Columbia University Libraries. This project is an extension of AlSum and Nelson's ``Thumbnail Summarization Techniques for Web Archives'' \cite{alsum-ecir14} and our previous work, funded by an incentive grant from Columbia University Libraries and the Andrew W. Mellon Foundation \cite{columbia-slides, columbia-arcvis-report}. 

\bibliographystyle{ACM-Reference-Format}
\bibliography{refs}


\begin{thebibliography}{27}


\ifx \showCODEN    \undefined \def \showCODEN     #1{\unskip}     \fi
\ifx \showDOI      \undefined \def \showDOI       #1{#1}\fi
\ifx \showISBNx    \undefined \def \showISBNx     #1{\unskip}     \fi
\ifx \showISBNxiii \undefined \def \showISBNxiii  #1{\unskip}     \fi
\ifx \showISSN     \undefined \def \showISSN      #1{\unskip}     \fi
\ifx \showLCCN     \undefined \def \showLCCN      #1{\unskip}     \fi
\ifx \shownote     \undefined \def \shownote      #1{#1}          \fi
\ifx \showarticletitle \undefined \def \showarticletitle #1{#1}   \fi
\ifx \showURL      \undefined \def \showURL       {\relax}        \fi
\providecommand\bibfield[2]{#2}
\providecommand\bibinfo[2]{#2}
\providecommand\natexlab[1]{#1}
\providecommand\showeprint[2][]{arXiv:#2}

\bibitem[\protect\citeauthoryear{Adar, Dontcheva, Fogarty, and Weld}{Adar
  et~al\mbox{.}}{2008}]%
        {adar-uist08}
\bibfield{author}{\bibinfo{person}{Eytan Adar}, \bibinfo{person}{Mira
  Dontcheva}, \bibinfo{person}{James Fogarty}, {and} \bibinfo{person}{Daniel~S.
  Weld}.} \bibinfo{year}{2008}\natexlab{}.
\newblock \showarticletitle{Zoetrope: Interacting with the ephemeral web}. In
  \bibinfo{booktitle}{\emph{Proceedings of the ACM Symposium on User Interface
  Software and Technology (UIST)}}. \bibinfo{pages}{239--248}.
\newblock


\bibitem[\protect\citeauthoryear{Alam and Nelson}{Alam and Nelson}{2016}]%
        {alam2016memgator}
\bibfield{author}{\bibinfo{person}{Sawood Alam} {and}
  \bibinfo{person}{Michael~L. Nelson}.} \bibinfo{year}{2016}\natexlab{}.
\newblock \showarticletitle{{MemGator} - A Portable Concurrent Memento
  Aggregator}. In \bibinfo{booktitle}{\emph{Proceedings of the 16th
  {ACM/IEEE-CS} Joint Conference on Digital Libraries}}.
  \bibinfo{pages}{243--244}.
\newblock


\bibitem[\protect\citeauthoryear{AlSum and Nelson}{AlSum and Nelson}{2014}]%
        {alsum-ecir14}
\bibfield{author}{\bibinfo{person}{Ahmed AlSum} {and}
  \bibinfo{person}{Michael~L. Nelson}.} \bibinfo{year}{2014}\natexlab{}.
\newblock \showarticletitle{Thumbnail Summarization Techniques for Web
  Archives}. In \bibinfo{booktitle}{\emph{Proceedings of the European
  Conference on Information Retrieval (ECIR)}}. \bibinfo{address}{Amsterdam},
  \bibinfo{pages}{299--310}.
\newblock


\bibitem[\protect\citeauthoryear{Arora, Li, Youtie, and Shapira}{Arora
  et~al\mbox{.}}{2016}]%
        {arora-15}
\bibfield{author}{\bibinfo{person}{Sanjay~K. Arora}, \bibinfo{person}{Yin Li},
  \bibinfo{person}{Jan Youtie}, {and} \bibinfo{person}{Philip Shapira}.}
  \bibinfo{year}{2016}\natexlab{}.
\newblock \showarticletitle{Using the {Wayback Machine} to mine websites in the
  social sciences: A methodological resource}.
\newblock \bibinfo{journal}{\emph{Journal of the Association for Information
  Science and Technology}} \bibinfo{volume}{67}, \bibinfo{number}{8}
  (\bibinfo{date}{Aug.} \bibinfo{year}{2016}), \bibinfo{pages}{1904--1915}.
\newblock


\bibitem[\protect\citeauthoryear{Balakireva, Shankar, Klein, Kremer, Powell,
  and {Van de Sompel}}{Balakireva et~al\mbox{.}}{2015}]%
        {timetravel}
\bibfield{author}{\bibinfo{person}{Luydmila Balakireva},
  \bibinfo{person}{Harihar Shankar}, \bibinfo{person}{Martin Klein},
  \bibinfo{person}{Ilya Kremer}, \bibinfo{person}{James Powell}, {and}
  \bibinfo{person}{Herbert {Van de Sompel}}.} \bibinfo{year}{2015}\natexlab{}.
\newblock \bibinfo{title}{{Time Travel APIs}}.
\newblock
  \bibinfo{howpublished}{\url{http://timetravel.mementoweb.org/guide/api/}}.
  (\bibinfo{year}{2015}).
\newblock


\bibitem[\protect\citeauthoryear{Charikar}{Charikar}{2002}]%
        {simhash-2002}
\bibfield{author}{\bibinfo{person}{Moses~S. Charikar}.}
  \bibinfo{year}{2002}\natexlab{}.
\newblock \showarticletitle{Similarity Estimation Techniques from Rounding
  Algorithms}. In \bibinfo{booktitle}{\emph{Proceedings of the Thirty-Fourth
  Annual ACM Symposium on Theory of Computing}}. \bibinfo{pages}{380--388}.
\newblock


\bibitem[\protect\citeauthoryear{Gunnam}{Gunnam}{2018}]%
        {gunnam-ms18}
\bibfield{author}{\bibinfo{person}{Maheedhar Gunnam}.}
  \bibinfo{year}{2018}\natexlab{}.
\newblock \bibinfo{title}{How {I} Changed Over Time: A webservice to summarize
  TimeMaps based on SimHashed HTML content}.
\newblock \bibinfo{howpublished}{ODU CS Masters Project Report}.
  (\bibinfo{date}{May} \bibinfo{year}{2018}).
\newblock
\newblock
\shownote{\url{http://www.cs.odu.edu/~mweigle/papers/gunnam-ms-proj-18.pdf}.}


\bibitem[\protect\citeauthoryear{Hamming}{Hamming}{1950}]%
        {hamming:1950}
\bibfield{author}{\bibinfo{person}{Richard~W. Hamming}.}
  \bibinfo{year}{1950}\natexlab{}.
\newblock \showarticletitle{Error detecting and error correcting codes}.
\newblock \bibinfo{journal}{\emph{The Bell System Technical Journal}}
  \bibinfo{volume}{29}, \bibinfo{number}{2} (\bibinfo{year}{1950}),
  \bibinfo{pages}{147--160}.
\newblock


\bibitem[\protect\citeauthoryear{Jones}{Jones}{2019}]%
        {raintale}
\bibfield{author}{\bibinfo{person}{Shawn~M. Jones}.}
  \bibinfo{year}{2019}\natexlab{}.
\newblock \bibinfo{title}{Raintale}.
\newblock
  \bibinfo{howpublished}{\url{https://ws-dl.blogspot.com/2019/07/2019-07-11-raintale-storytelling-tool.html},
  \url{https://github.com/oduwsdl/raintale}}.   (\bibinfo{year}{2019}).
\newblock


\bibitem[\protect\citeauthoryear{Jones, {Van de Sompel}, Shankar, Klein, Tobin,
  and Grover}{Jones et~al\mbox{.}}{2016}]%
        {jones-plosone2016}
\bibfield{author}{\bibinfo{person}{Shawn~M. Jones}, \bibinfo{person}{Herbert
  {Van de Sompel}}, \bibinfo{person}{Harihar Shankar}, \bibinfo{person}{Martin
  Klein}, \bibinfo{person}{Richard Tobin}, {and} \bibinfo{person}{Claire
  Grover}.} \bibinfo{year}{2016}\natexlab{}.
\newblock \showarticletitle{Scholarly Context Adrift: Three out of Four {URI}
  References Lead to Changed Content}.
\newblock \bibinfo{journal}{\emph{{PLoS ONE}}} \bibinfo{volume}{11},
  \bibinfo{number}{12} (\bibinfo{year}{2016}).
\newblock


\bibitem[\protect\citeauthoryear{Kelly}{Kelly}{2017}]%
        {thumbnail-code}
\bibfield{author}{\bibinfo{person}{Mat Kelly}.}
  \bibinfo{year}{2017}\natexlab{}.
\newblock \bibinfo{title}{{ArchiveThumbnails}}.
\newblock
  \bibinfo{howpublished}{\url{https://github.com/machawk1/ArchiveThumbnails}}.
   (\bibinfo{year}{2017}).
\newblock


\bibitem[\protect\citeauthoryear{Kelly, Nelson, and Weigle}{Kelly
  et~al\mbox{.}}{2015}]%
        {columbia-slides}
\bibfield{author}{\bibinfo{person}{Mat Kelly}, \bibinfo{person}{Michael~L.
  Nelson}, {and} \bibinfo{person}{Michele~C. Weigle}.}
  \bibinfo{year}{2015}\natexlab{}.
\newblock \bibinfo{title}{Visualizing Digital Collections of Web Archives
  (slides)}.
\newblock
  \bibinfo{howpublished}{\url{http://www.slideshare.net/matkelly01/visualizing-digital-collections-of-web-archives-from-columbia-web-archiving-collaboration-conference}}.
    (\bibinfo{date}{June} \bibinfo{year}{2015}).
\newblock


\bibitem[\protect\citeauthoryear{Klein and Nelson}{Klein and Nelson}{2008}]%
        {lexsig:ecdl08}
\bibfield{author}{\bibinfo{person}{Martin Klein} {and}
  \bibinfo{person}{Michael~L. Nelson}.} \bibinfo{year}{2008}\natexlab{}.
\newblock \showarticletitle{Revisiting Lexical Signatures to (Re-)Discover Web
  Pages}. In \bibinfo{booktitle}{\emph{Proceedings of the 12th European
  Conference on Research and Advanced Technology for Digital Libraries}}.
  \bibinfo{pages}{371--382}.
\newblock


\bibitem[\protect\citeauthoryear{Klein, {Van de Sompel}, Sanderson, Shankar,
  Balakireva, Zhou, and Tobin}{Klein et~al\mbox{.}}{2014}]%
        {klein-plosone2014}
\bibfield{author}{\bibinfo{person}{Martin Klein}, \bibinfo{person}{Herbert {Van
  de Sompel}}, \bibinfo{person}{Robert Sanderson}, \bibinfo{person}{Harihar
  Shankar}, \bibinfo{person}{Lyudmila Balakireva}, \bibinfo{person}{Ke Zhou},
  {and} \bibinfo{person}{Richard Tobin}.} \bibinfo{year}{2014}\natexlab{}.
\newblock \showarticletitle{Scholarly Context Not Found: One in Five Articles
  Suffers from Reference Rot}.
\newblock \bibinfo{journal}{\emph{{PLoS ONE}}} \bibinfo{volume}{9},
  \bibinfo{number}{12} (\bibinfo{year}{2014}).
\newblock


\bibitem[\protect\citeauthoryear{Lim, Wang, Padmanabhan, Vitter, and
  Agarwal}{Lim et~al\mbox{.}}{2001}]%
        {lim2001characterizing}
\bibfield{author}{\bibinfo{person}{L. Lim}, \bibinfo{person}{M. Wang},
  \bibinfo{person}{S. Padmanabhan}, \bibinfo{person}{J. Vitter}, {and}
  \bibinfo{person}{R. Agarwal}.} \bibinfo{year}{2001}\natexlab{}.
\newblock \showarticletitle{Characterizing web document change}.
\newblock \bibinfo{journal}{\emph{Advances in Web-Age Information Management}}
  (\bibinfo{year}{2001}), \bibinfo{pages}{133--144}.
\newblock


\bibitem[\protect\citeauthoryear{Mayer and Moreno}{Mayer and Moreno}{2003}]%
        {Mayer2003}
\bibfield{author}{\bibinfo{person}{Richard~E Mayer} {and}
  \bibinfo{person}{Roxana Moreno}.} \bibinfo{year}{2003}\natexlab{}.
\newblock \showarticletitle{{Nine ways to reduce cognitive load in multimedia
  learning}}.
\newblock \bibinfo{journal}{\emph{Educational Psychologist}}
  \bibinfo{volume}{38}, \bibinfo{number}{1} (\bibinfo{year}{2003}),
  \bibinfo{pages}{43--52}.
\newblock
\showISSN{0046-1520}


\bibitem[\protect\citeauthoryear{Milligan}{Milligan}{2016}]%
        {milligan16}
\bibfield{author}{\bibinfo{person}{Ian Milligan}.}
  \bibinfo{year}{2016}\natexlab{}.
\newblock \showarticletitle{Lost in the Infinite Archive: The Promise and
  Pitfalls of Web Archives}.
\newblock \bibinfo{journal}{\emph{International Journal of Humanities and Arts
  Computing}} \bibinfo{volume}{10}, \bibinfo{number}{1} (\bibinfo{year}{2016}),
  \bibinfo{pages}{78--94}.
\newblock


\bibitem[\protect\citeauthoryear{Padia, AlNoamany, and Weigle}{Padia
  et~al\mbox{.}}{2012}]%
        {padia-jcdl12}
\bibfield{author}{\bibinfo{person}{Kalpesh Padia}, \bibinfo{person}{Yasmin
  AlNoamany}, {and} \bibinfo{person}{Michele~C. Weigle}.}
  \bibinfo{year}{2012}\natexlab{}.
\newblock \showarticletitle{Visualizing Digital Collections at {Archive-It}}.
  In \bibinfo{booktitle}{\emph{Proceedings of the ACM/IEEE Joint Conference on
  Digital Libraries (JCDL)}}. \bibinfo{address}{Washington, DC},
  \bibinfo{pages}{15--18}.
\newblock


\bibitem[\protect\citeauthoryear{Shankar}{Shankar}{2017}]%
        {shankar-ms17}
\bibfield{author}{\bibinfo{person}{Surbhi Shankar}.}
  \bibinfo{year}{2017}\natexlab{}.
\newblock \bibinfo{title}{Visualizing Thumbnails Of Archived Web Pages}.
\newblock \bibinfo{howpublished}{ODU CS Masters Project Report}.
  (\bibinfo{date}{May} \bibinfo{year}{2017}).
\newblock
\newblock
\shownote{\url{http://www.cs.odu.edu/~mweigle/papers/shankar-ms-proj-17.pdf}.}


\bibitem[\protect\citeauthoryear{Shaw}{Shaw}{2011}]%
        {timeline-propublica}
\bibfield{author}{\bibinfo{person}{Al Shaw}.} \bibinfo{year}{2011}\natexlab{}.
\newblock \bibinfo{title}{{TimelineSetter}: Easy Timelines From Spreadsheets,
  Now Open to All}.
\newblock
  \bibinfo{howpublished}{\url{https://www.propublica.org/nerds/timelinesetter-easy-timelines-from-spreadsheets-now-open-to-all}}.
    (\bibinfo{date}{April} \bibinfo{year}{2011}).
\newblock


\bibitem[\protect\citeauthoryear{Sherratt}{Sherratt}{2020}]%
        {glam-workbench}
\bibfield{author}{\bibinfo{person}{Tim Sherratt}.}
  \bibinfo{year}{2020}\natexlab{}.
\newblock \bibinfo{title}{{GLAM Workbench}: Web Archives}.
\newblock
  \bibinfo{howpublished}{\url{https://glam-workbench.github.io/web-archives/}}.
    (\bibinfo{year}{2020}).
\newblock


\bibitem[\protect\citeauthoryear{Starbird}{Starbird}{2016}]%
        {starbird-blog16}
\bibfield{author}{\bibinfo{person}{Kate Starbird}.}
  \bibinfo{year}{2016}\natexlab{}.
\newblock \bibinfo{title}{Tracing Disinformation Trajectories from the 2010
  {Deepwater Horizon} Oil Spill}.
\newblock
  \bibinfo{howpublished}{\url{https://medium.com/hci-design-at-uw/tracing-disinformation-trajectories-from-the-2010-deepwater-horizon-oil-spill-79e8116e08f4}}.
    (\bibinfo{date}{Dec.} \bibinfo{year}{2016}).
\newblock


\bibitem[\protect\citeauthoryear{Starbird, Dailey, Walker, Leschine, Pavia, and
  Bostrom}{Starbird et~al\mbox{.}}{2015}]%
        {starbird-2015}
\bibfield{author}{\bibinfo{person}{Kate Starbird}, \bibinfo{person}{Dharma
  Dailey}, \bibinfo{person}{Ann~Hayward Walker}, \bibinfo{person}{Thomas~M.
  Leschine}, \bibinfo{person}{Robert Pavia}, {and} \bibinfo{person}{Ann
  Bostrom}.} \bibinfo{year}{2015}\natexlab{}.
\newblock \showarticletitle{Social Media, Public Participation, and the 2010
  {BP Deepwater Horizon} Oil Spill}.
\newblock \bibinfo{journal}{\emph{Human and Ecological Risk Assessment: An
  International Journal}} \bibinfo{volume}{21}, \bibinfo{number}{3}
  (\bibinfo{year}{2015}), \bibinfo{pages}{605--630}.
\newblock


\bibitem[\protect\citeauthoryear{{Van de Sompel}, Nelson, and Sanderson}{{Van
  de Sompel} et~al\mbox{.}}{2013}]%
        {memento-rfc}
\bibfield{author}{\bibinfo{person}{Herbert {Van de Sompel}},
  \bibinfo{person}{Michael~L. Nelson}, {and} \bibinfo{person}{Robert
  Sanderson}.} \bibinfo{year}{2013}\natexlab{}.
\newblock \bibinfo{title}{{HTTP} Framework for Time-Based Access to Resource
  States -- Memento}.
\newblock   (\bibinfo{year}{2013}).
\newblock
\newblock
\shownote{RFC 7089.}


\bibitem[\protect\citeauthoryear{Weigle}{Weigle}{2017}]%
        {weigle-neh-2017}
\bibfield{author}{\bibinfo{person}{Michele~C. Weigle}.}
  \bibinfo{year}{2017}\natexlab{}.
\newblock \bibinfo{title}{{Visualizing Webpage Changes Over Time - new NEH
  Digital Humanities Advancement Grant}}.
\newblock
  \bibinfo{howpublished}{\url{https://ws-dl.blogspot.com/2017/10/2017-10-16-visualizing-webpage-changes.html}}.
    (\bibinfo{year}{2017}).
\newblock


\bibitem[\protect\citeauthoryear{Weigle and Nelson}{Weigle and Nelson}{2015}]%
        {columbia-arcvis-report}
\bibfield{author}{\bibinfo{person}{Michele~C. Weigle} {and}
  \bibinfo{person}{Michael~L. Nelson}.} \bibinfo{year}{2015}\natexlab{}.
\newblock \bibinfo{title}{Visualizing Digital Collections of Web Archives -
  Final Report for {Columbia Libraries Web Archiving Incentive Program}}.
\newblock
  \bibinfo{howpublished}{\url{http://www.cs.odu.edu/~mweigle/papers/Columbia-arcvis-final-report-2015.pdf}}.
    (\bibinfo{date}{June} \bibinfo{year}{2015}).
\newblock


\bibitem[\protect\citeauthoryear{West and Franko}{West and Franko}{2018}]%
        {gifshot}
\bibfield{author}{\bibinfo{person}{Chase West} {and} \bibinfo{person}{Greg
  Franko}.} \bibinfo{year}{2018}\natexlab{}.
\newblock \bibinfo{title}{GifShot}.
\newblock \bibinfo{howpublished}{\url{https://yahoo.github.io/gifshot/}}.
  (\bibinfo{year}{2018}).
\newblock


\end{thebibliography}

\end{document}